# Mobility engineering and metal-insulator transition in monolayer MoS$_2$

B. Radisavljevic and A. Kis[*]

*Electrical Engineering Institute, Ecole Polytechnique Federale de Lausanne (EPFL), CH-1015 Lausanne, Switzerland*

*Correspondence should be addressed to: Andras Kis, andras.kis@epfl.ch

**Two-dimensional (2D) materials are a new class of materials with interesting physical properties and ranging from nanoelectronics to sensing and photonics. In addition to graphene, the most studied 2D material, monolayers of other layered materials such as semiconducting dichalcogenides MoS$_2$ or WSe$_2$ are gaining in importance as promising insulators and channel materials for field-effect transistors (FETs). The presence of a direct band gap in monolayer MoS$_2$ due to quantum mechanical confinement, allows room-temperature field-effect transistors with an on/off ratio exceeding $10^8$. The presence of high-k dielectrics in these devices enhanced their mobility, but the mechanisms are not well understood. Here, we report on electrical transport measurements on MoS$_2$ FETs in different dielectric configurations. Mobility dependence on temperature shows clear evidence of the strong suppression of charge impurity scattering in dual-gate devices with a top-gate dielectric together with phonon scattering that shows a weaker than expected temperature dependence. High levels of doping achieved in dual-gate devices also**

**allow the observation of a metal-insulator transition in monolayer $MoS_2$. Our work opens up the way to further improvements in 2D semiconductor performance and introduces $MoS_2$ as an interesting system for studying correlation effects in mesoscopic systems.**

Molybdenum disulphide ($MoS_2$) is a typical layered transition-metal dichalcogenide (TMD) semiconductor[1] with potential applications that could complement those of graphene. Because neighboring layers in TMD crystals are weakly bound via van der Waals interaction, single atomic crystals composed of one or several layers can be extracted using either the micromechanical cleavage technique[2] originally developed for the production of graphene or liquid phase exfoliation.[3,4] Few-layer large-area $MoS_2$ can also be grown using CVD-like growth techniques.[5,6] The strong covalent bonding between metal and chalcogenide atoms results in a high mechanical strength[7] of $MoS_2$ membranes[8] and electrical breakdown current densities at least 50 times higher than in copper.[9] In contrast to graphene, the presence of a band gap in monolayer $MoS_2$ and other semiconducting dichalcogenides allows the fabrication of transistors that can be turned off and used as switches.[10] These transistors showed a current ON/OFF ratio ~$10^8$, low subthreshold swing (74 mV/dec) and negligible OFF current (25 fA/μm).[10] Logic circuits[11] and amplifiers[12] with high gain based on monolayer $MoS_2$ and ring-oscillators[13] based on bilayer $MoS_2$ have also been demonstrated.

Transistors based on monolayer $WSe_2$ (ref 14) as well as thin multilayer $WS_2$ and $MoSe_2$ have also been recently demonstrated[15,16] while superconductivity in 20-nm thick $MoS_2$ was achieved at high electron concentrations using ionic-liquid gating.[17]

Monolayer MoS$_2$ has electronic and optical properties that are fundamentally different from those of thicker layers due to quantum-mechanical confinement.[18,19] Whereas bulk MoS$_2$ is indirect gap semiconductor band gap of 1.2 eV, single-layer MoS$_2$ is a direct gap semiconductor.[18-21] The lack of inversion symmetry results in strong coupling of spin and valley degrees of freedom that can be detected using circularly polarized light[22-24] and could be used in novel devices based on valley Hall effect.[25] The atomic scale thickness (6.5Å) of monolayer MoS$_2$, smaller than the screening length also allows a large degree of electrostatic control over the electrical conductivity. Together with the absence of dangling bonds, this would allow transistors based on monolayer MoS$_2$ to outperform silicon transistors at the scaling limit.[26,27]

Previous measurements have shown that the room-temperature mobility of bulk MoS$_2$ is in the 200-500 cm$^2$/Vs range and is limited by phonon scattering.[28] Exfoliation of single layers onto SiO$_2$ results in a decrease of mobility down to the 0.1-10 cm$^2$/Vs range[2,10] while charge traps[29] present at the interface between the substrate and the MoS$_2$ layer have recently been proposed as the dominant cause for such low room temperature mobility in MoS$_2$ devices. Understanding the origin of this mobility degradation and finding a way to restore the mobility to bulk values or even further enhance it would allow us to unlock the full technological potential of this material.

The encapsulation of monolayer MoS$_2$ in a high-κ dielectric environment[30] was shown to result in an increase of the room-temperature mobility.[10] This was tentatively assigned to reduced Coulomb scattering due to the high-κ dielectric environment[30] and possible modification of phonon dispersion in MoS$_2$ monolayers. An increase of mobility with the dielectric deposition, similar to that in monolayers was also observed in multilayer samples[31,32] and monolayer samples with polymer gating.[33] Previous mobility estimates for monolayer MoS$_2$ are however based on two-contact measurements and lack the information on their temperature dependence. More accurate measurements are needed in order to gain better understanding of the various mechanisms that could limit the mobility in monolayer MoS$_2$.

Temperature-dependent measurements of mobility could be used to distinguish between different mechanisms limiting the mobility in monolayer MoS$_2$ and their relative contributions. In the phonon-limited high-temperature part, the mobility is expected to follow a $\mu \sim T^{-\gamma}$ temperature dependence with $\gamma = 1.69$ and mobility reaching a room-temperature value ~410 cm$^2$/Vs according to first-principle calculations by Kaasbjerg et al.[34] The deposition of a top-gate dielectric is expected to mechanically quench the homopolar phonon mode and reduce the coefficient $\gamma$ to 1.52. Measurements on bulk crystals show $\gamma$ of 2.6.[28]

Here, we report on mobility measurements in monolayer MoS$_2$ based on the Hall effect. This allows us to remove the effect of contact resistance and also directly measure the gate-modulated charge density and gate capacitance necessary for the accurate measurements of the field-effect mobility. Our devices are field-effect transistors in single-gate and dual-gate configurations as shown on Figure 1. Degenerately-doped Si wafers covered with 270 nm thermally grown SiO$_2$ serve as the substrate and back-gate for the MoS$_2$ devices. Single and few-layer MoS$_2$ flakes are obtained by standard micromechanical cleavage[2] technique. Flakes are identified by an optical microscope and their thickness is ascertained by optical contrast measurements[35] and atomic-force microscopy. The source, drain and voltage probes were defined by electron-beam lithography followed by deposition of 90 nm thick Au electrodes. In the inset of Figure 1a we present an optical image of the device after standard lift-off procedure performed in acetone. In order to remove resist residue and decrease contact resistance in our devices, we perform annealing at 200 ºC in Ar atmosphere for 2 hours. After this step, we shape the $MoS_2$ flakes into Hall bars using oxygen plasma etching and an e-beam defined etching mask. Some of our devices were further processed and a 30 nm-thick $HfO_2$ layer was deposited by atomic layer deposition (ALD) followed by another e-beam lithography process defining top-gate electrode. Top-gate electrode is made by depositing Cr/Au (10/50 nm) layer by electron-beam evaporation and lift-off in acetone. The optical image of one of our top-gated devices is shown in Figure 1a. All devices are wirebonded onto chip carriers and transferred to a cryostat where the transport measurements were performed in vacuum from room temperature down to 300 mK.

We have performed measurements on two devices in single-gate configuration, two devices in single-gate configuration covered with a 30 nm thick $HfO_2$ layer and six devices in dual-gate configuration (Supplementary Table 1). In the case of single-gate devices, conductance defined as $G = I_{ds}/(V_1-V_2)$ is measured while sweeping the back-gate voltage $V_{bg}$, Figure 1b (upper schematic) and maintaining the drain-source bias $V_{ds}$ constant. During the characterization of dual-gate devices, we maintain the back-gate grounded and sweep the top-gate voltage $V_{tg}$, Figure 1b (lower schematic). By using the top-gate we can induce stronger electrostatic doping of our monolayer $MoS_2$ owing to the higher dielectric constant and smaller thickness of the $HfO_2$ layer ($\varepsilon_{r2}$ ~ 19, $d_{ox2}$ ($HfO_2$) = 30 nm) compared to the bottom-gate $SiO_2$ ($\varepsilon_{r1}$ ~ 3.9, $d_{ox1}$ ($SiO_2$) = 270 nm).

Conductance measurements are performed in the four-probe configuration for all devices presented here. All our devices show behavior typical of n-type semiconductors. A typical conductance $G$ dependence on the gate voltage for a single-gate device is shown in Figure 2a, measured up to the back-gate voltage $V_{bg}$ = 40 V that corresponds to a charge concentration of $n_{2D}$ ~ 3.6 · $10^{12}$ $cm^{-2}$ calculated using the parallel-plate capacitor model, with $n_{2D} = C_{ox1} \Delta V_{bg}/e$, where $C_{ox1} = \varepsilon_0 \varepsilon_{r1}/d_{ox1}$, $\varepsilon_0$ = 8.85×$10^{-12}$ F/m, $\Delta V_{bg} = V_{bg} - V_{bg,th}$. The value of threshold voltage $V_{bg,th}$ varies for each device and is close to its pinch-off voltage estimated from the conductance curves. We find that temperature variation of the conductance $G$ in a single-gate monolayer device (Figure 2b), in the high-temperature regime (80 K $\leq T \leq$ 280 K), can be modeled with thermally-activated transport where the conductance is described by expression:

$$G = G_0(T) e^{-E_a/k_B T}$$

where $E_a$ is the activation energy, $k_B$ the Boltzmann constant and $G_o(T)$ the temperature-dependent parameter extracted from the fitting curves. Good agreement of the data to activation transport model at higher temperatures is suggestive of charge transport in a two-dimensional system that is thermally activated. At temperatures $T \leq$ 80 K we observe that the variation of $G$ weakens for almost all $V_{bg}$ values. That can be explained with the fact that at lower temperatures hopping through localized states becomes dominant[29] and the system is driven into a strongly localized regime.

In Figure 2c we show the temperature dependence of the mobility in this device. Mobility is extracted from the conductance curves in the 30 - 40 V range of back-gate voltage $V_{bg}$, using the expression for field-effect mobility $\mu = \left[ dG / dV_{bg} \right] \times \left[ L_{12} / (WC_{ox1}) \right]$. The temperature dependence is characterized by a peak at ~200K. Below 200 K, we observe a decrease of the mobility as the temperature is lowered down to 4 K. This behavior is consistent with mobility limited by scattering from charged impurities.[36] Increasing the temperature above 200 K, also results in a strong decrease of the mobility from the peak value of 18 cm$^2$/Vs, related to electron-phonon scattering that becomes the dominant mechanism at higher temperatures.[34] We fit this part of the curve with generic temperature dependence of the mobility $\mu \sim T^{-\gamma}$, where the exponent $\gamma$ depends on the dominant phonon scattering mechanism. From the fit we find the value of $\gamma \approx 1.4$, in good agreement with a theoretical predictions for monolayer MoS$_2$ ($\gamma \approx 1.69$).[34]

We now turn to dual-gated devices, with a typical top-gating dependence of the four-contact conductance given in Figure 3a. The use of the top-gate allows higher degree of electrostatic control and doping over $n_{2D} \sim 3 \cdot 10^{13}$ cm$^{-2}$, typical for single-gated devices. We observe here an insulating behavior that persists until the top-gate voltage of $V_{tg}$ = 2.2 V. At this point, corresponding to a charge concentration $n_{2D} \sim 1 \cdot 10^{13}$ cm$^{-2}$ (as measured from Hall-effect), monolayer MoS$_2$ enters a metallic state and the associated metal-insulator transition (MIT)[37] is observed, the first of its kind in a two-dimensional semiconductor, Figure 3a. Figure 3b shows the temperature dependence of the device conductance for different values of the charge density $n_{2D}$ and a metal-insulator transition. This striking feature occurs when the resistivity is of the order of the quantum resistance $h/e^2$. This point is called the minimum of metallic conductivity and for a long time considered not to exist in two-dimensional electronic systems.[38] The first step in our analysis is to define the critical point of the MIT. Inspecting the Figure 3a, we can see that each two consecutive isotherms of $G$ ($V_{tg}$) cross each other at some value of $V_{tg}$. As it can be clearly seen, these intersections are temperature dependent, and an unambiguous determination of the transition is therefore not possible. Fortunately, at the temperatures lower than 80 K, the crossing point appears to be independent of the temperature and emerges at a well-defined point $V_{tg}$ = 2.2 V, which clearly separates the metallic and insulating phases. This transition point is the direct consequence of quantum interference effects of weak and strong localization. At lower carrier concentrations (< $n_{2D}$ ~ 1 × 10$^{13}$ cm$^{-2}$) system is in the insulating state and strong localization[39] prevails. This charge density is comparable to that recorded for 20-nm thick MoS$_2$.[17] As the top-gate bias is increased above $V_{tg}$ = 2.2 V (concentration above $n$ ~1 × 10$^{13}$ cm$^{-2}$), the system is driven into a metallic phase and weak localization appears to be the dominant effect. In this regime, due to high carrier concentration, Coulomb electron-electron interaction plays an important role as well.

We can now investigate the Ioffe-Regel criterion[40-42] for two-dimensional semiconductors which predicts the existence of a metal-insulator transition when the parameter $k_F \cdot l_e$ satisfies the criterion $k_F \cdot l_e$ ~ 1, with the Fermi wave vector $k_F = \sqrt{2\pi n_{2D}}$, and mean free path of electrons $l_e = \hbar k_F \sigma / n_{2D} e^2$, where σ is the sheet conductivity $\sigma = G L_{12} / W$ with $L_{12}$ = 1.55 μm and $W$ = 1.9 μm the distance between voltage probes and device width, respectively. According to this criterion, for $k_F \cdot l_e \gg 1$ the phase is metallic while for $k_F \cdot l_e \ll 1$, the phase is insulating. For our device, at the crossing point of $V_{tg}$ = 2.2 V, we have $k_F \cdot l_e$ ~ 2.5, in good agreement with the theory. Our other devices also display $k_F \cdot l_e$ close to 2, Supplementary Table 1.

Temperature dependence of mobility is extracted from conductance curves in the $V_{tg}$ = 2-5 V range that can be considered as linear regime for all temperatures, using the expression for field-effect mobility $\mu = \left[ dG/dV_{bg} \right] \times \left[ L_{12} / \left( W C_{tg,Hall} \right) \right]$ with capacitance $C_{tg,Hall}$ extracted from Hall-effect measurements. For all monolayer, double and three-layer dual-gate devices that we characterized, we observe monotonous increase of the mobility as the temperature is decreased with a saturation at low temperatures. Figure 3c shows the temperature dependence of mobility for the main device presented here. The mobility at 4K is 168 cm$^2$/Vs, reaching 60 cm$^2$/Vs at 240 K. This makes a distinct difference from devices fabricated in single-gate configuration where the monotonous decrease of the mobility is observed as the temperature is lowered from 250 K down to 4 K (Figure 2c). We relate this behavior to effective damping of static Coulomb scattering on charge impurities due to the presence of the high-k dielectric and the metallic top-gate that changes the dielectric environment of monolayer MoS$_2$.[27] In the phonon-limited part, the mobility can be fitted to the expression $\mu \sim T^{-\gamma}$, with the exponent $\gamma \approx 0.73$ in the 100-300K range (black solid line in Figure 3c). For all our double-gated monolayer devices we find this exponent to be between 0.3 and 0.75, while for one double-layer device we find a value of 1.47. These values for monolayer MoS$_2$ much smaller than the theoretically predicted value of $\gamma \approx 1.52$ (ref [34]) or bulk crystals ($\gamma \approx 2.6$, ref [28]). This indicates that in addition to the quenching of the homopolar phonon mode other mechanisms might influence the mobility of monolayer MoS$_2$ in dual-gated devices, for example phonon screening induced by the metallic top gate or a change in the strength of electron-phonon coupling. Further theoretical modeling could shed more light on these mechanisms.

Just as in the case of single-gated devices, we model the temperature dependence of the conductance $G$ in the insulating regime of our double-gated devices with thermally activated behavior, Figure 4a. Here, we can observe that the activated behavior fits are data very well in the 100-250K temperature range. Extracted activation energies $E_a$ are shown in Figure 4b.

We have performed Hall effect measurements on all $MoS_2$ devices covered with a dielectric layer presented here in order to accurately determine the mobility, density of charge carriers and the capacitive coupling of $MoS_2$ layers to control gate electrodes (bottom or top gates). Figure 5a shows the transverse Hall resistance $R_{xy}$ of our main dual-gated monolayer device which follows a linear dependence on the magnetic field $B$ for different values of top-gate voltages $V_{tg}$. From the inverse slope of $R_{xy}$ we can directly determine the electron density $n_{2D}$ in the $MoS_2$ channel. The variation of the electron density extracted from $R_{xy}$ as a function of the top-gate voltage $V_{tg}$ is shown on Figure 5b. The slope of this dependence gives directly the capacitance $C_{tg,Hall} = 3.17 \cdot 10^{-7}$ F/cm$^2$ used in calculation of the field-effect mobility (Figures 3c). We also directly measure the capacitive coupling between the channel and the bottom gate in devices where the $MoS_2$ channel is covered with a dielectric layer and in devices with disconnected top gates and compare them to the geometric capacitance per unit area calculated using the parallel-plate capacitance model $C_{geom} = \varepsilon_0 \varepsilon_r / d_{ox,bottom}$ where $d_{ox,bottom}$ is the thickness of the bottom-gate oxide.[10,43] We find that encapsulation in a dielectric can increase the capacitive coupling from $C_{geom}$ by a factor of 2.4, similarly to previous reports on graphene devices[44] while disconnecting the top-gate dielectric increases the capacitive coupling by a factor of 53 (Suplementary Fig S1). These measurements prove that the capacitance can be underestimated in a complicated dielectric environment, both in the case of disconnected top gates[10,43] and encapsulation[45] resulting in mobility values that are likely to be overestimated. In order to accurately measure the field-effect mobility of FETs based on 2D materials one needs to measure the actual capacitance using either CV[32] or Hall effect measurements as outlined here.

In conclusion, we have performed conductance and mobility measurements on monolayer $MoS_2$ field-effect transistors in single-gate and dual-gate configuration. Using a top-gate in the dual-gate configuration and solid-state dielectrics, we were able to tune the charge carrier density over $n \sim 4 \times 10^{13}$ cm$^{-2}$ inducing the transition from insulating to metal phase in monolayer $MoS_2$. This transition point is in good agreement with theory and shows that monolayer $MoS_2$ could be an interesting new material system for investigating low-dimensional correlated electron behaviour. The metal-insulator transition could also be exploited to realize new types of switches, especially fast optoelectronic switches based on differences in optical transmission in metallic and insulating states.[46] In addition to allowing high charge densities, the high-k $HfO_2$ used as a top-gate dielectric also changes the dielectric environment and effectively screens Coulomb scattering which results in mobility improvement in dual-gate devices. Additionally, the presence of the top gate dielectric and metal electrode results in a quenching of the homopolar mode which is polarized in the direction normal to the layer, leading to a strong decrease of the mobility exponent $\gamma$ in $\mu \sim T^{-\gamma}$. Our results provide a new picture of the mobility issue in different configurations of $MoS_2$ devices, which should shed new light on the directions for further improvements in device quality and characterization techniques.

**METHODS**

MoS$_2$ flakes are exfoliated from molybdenite crystals (SPI Supplies Brand Moly Disulfide) by scotch-tape micromechanical cleavage technique. ALD is performed in a Beneq system and in an home-built ALD reactor using a reaction of H$_2$O with tetrakis(ethyl-methylamido)hafnium. Electrical characterization is carried out using National Instruments DAQ cards, SR570 current preamplifiers, SR560 low noise voltage preamplifiers, and an Oxford Instruments Heliox cryo-magnetic system.

**ACKNOWLEDGMENTS**

We thank Walter Escoffier (LNCMI CNRS), Bertrand Raquet (LNCMI CNRS) and Simone Bertolazzi (EPFL) for useful discussions as well as Jean-Savin Heron (EPFL) for technical support. Device fabrication was carried out in part in the EPFL Center for Micro/Nanotechnology (CMI). We thank Zdenek Benes (CMI) for technical support with e-beam lithography and Aleksandra Radenovic and Michael Whitwick (EPFL) for support with ALD deposition. This work was financially supported by ERC grant no. 240076, FLATRONICS: Electronic devices based on nanolayers.

# FIGURES

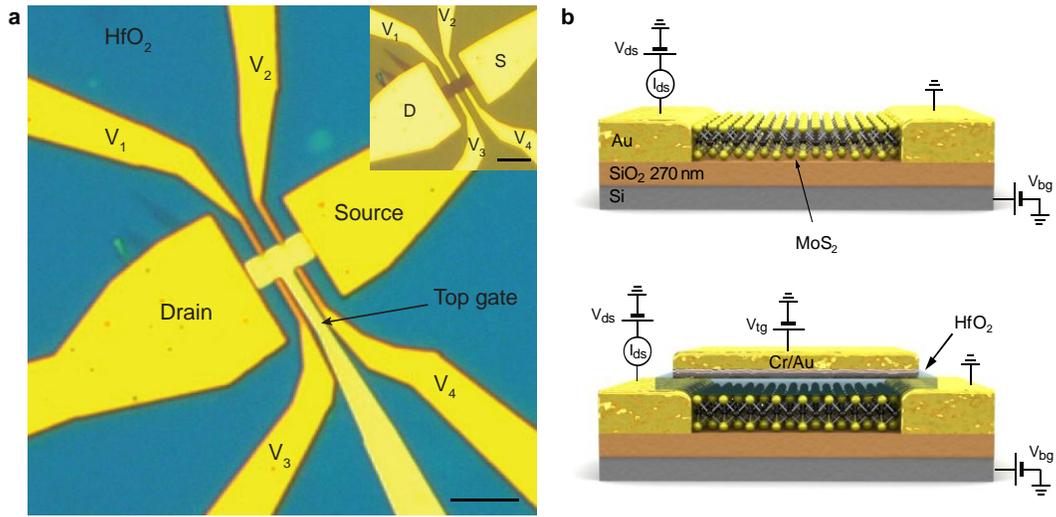

**Figure 1. Fabrication of single-gated and dual-gated MoS$_2$ devices**. **a**, Optical image of the MoS$_2$ dual-gated device used in our measurements. Inset shows the single-gate version of the same device before ALD deposition of HfO$_2$ and top-gate electrode fabrication. **b**, Cross-sectional views of devices based on single-layer MoS$_2$ in a single-gate (upper schematic) and dual-gate (lower schematic) configuration. Gold leads are used for the source, drain and voltage-probes ($V_1$, $V_2$, $V_3$ and $V_4$). Voltage probes have been omitted from the drawing. The silicon substrate, covered with a 270 nm thick SiO$_2$ layer was used as the back-gate. The top-gate dielectric was a 30 nm thick HfO$_2$ layer. Scale bars are 5 μm long.

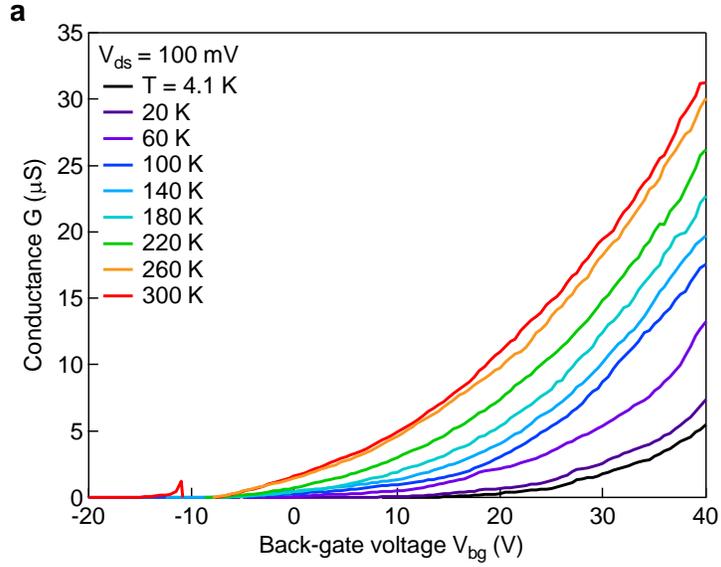

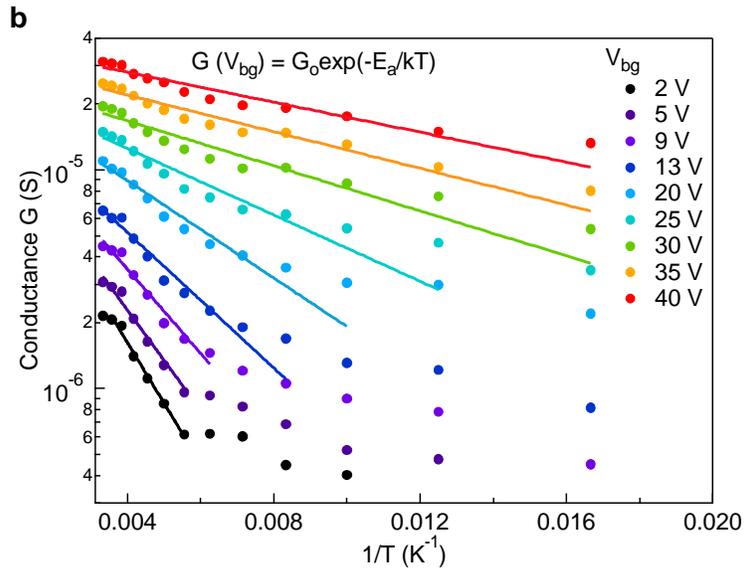

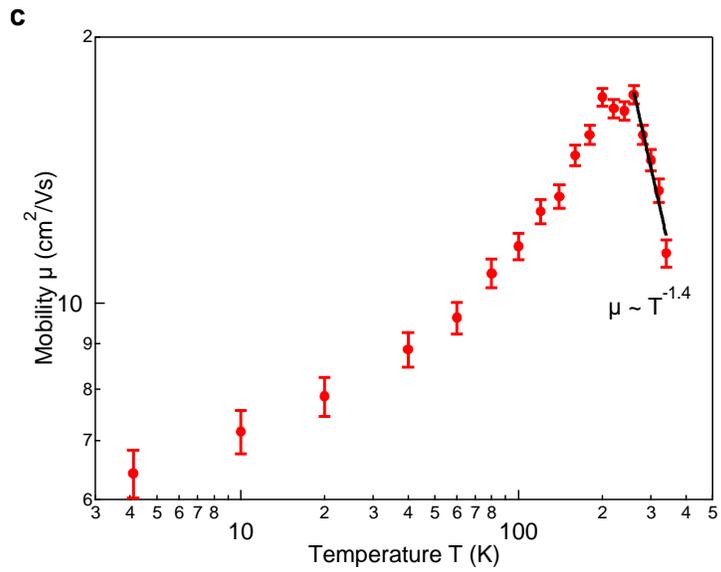

**Figure 2. Electron transport in single-gate monolayer MoS$_2$ supported on SiO$_2$. a**, Conductance $G$ as a function of back-gate voltage $V_{bg}$ for a single-gate monolayer MoS$_2$ device acquired at different temperatures. **b**, Arrhenius plot of the conductance $G$ for different values of the back-gate voltage. Solid lines are linear fits to the data showing activated behavior for limited regions of temperature and back-gate voltage (charge density). **c**, The dependence of the mobility on temperature shows a pronounced low-temperature regime consistent with transport dominated by scattering from charged impurities. Above ∼200 K, the mobility is limited by phonon scattering and follows a $\mu \sim T^{-1.4}$ dependence.

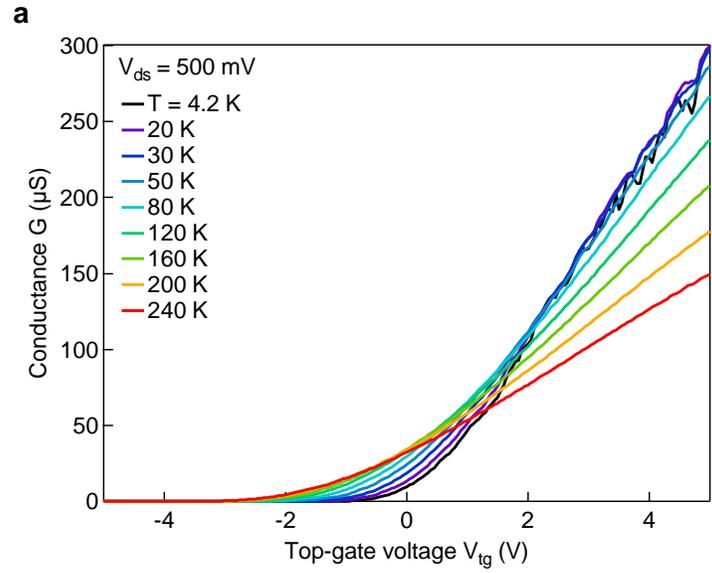

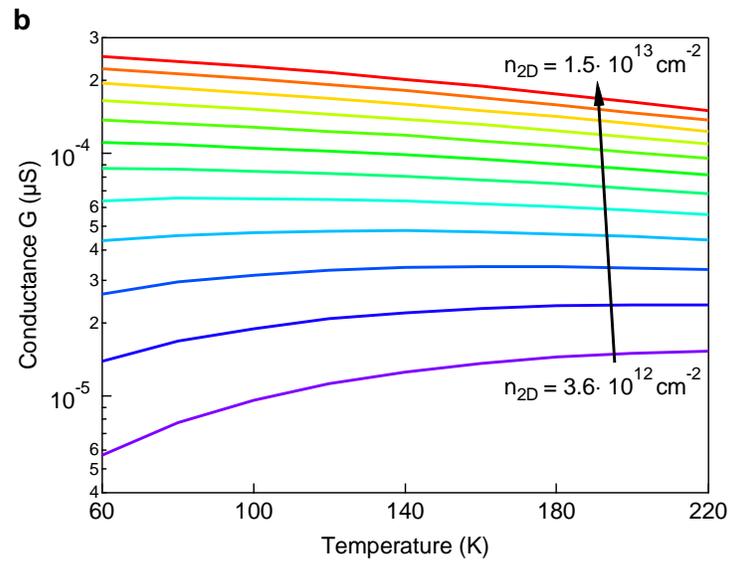

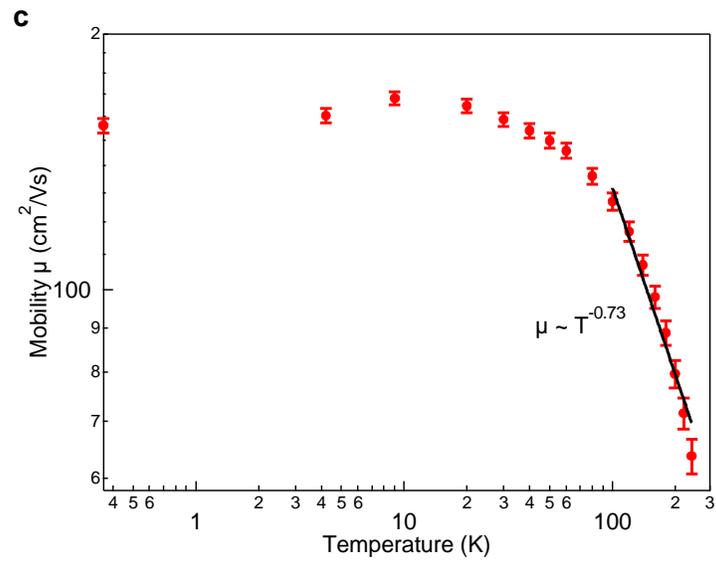
19

**Figure 3. Electron transport in dual-gated monolayer MoS$_2$. a,** Conductance $G$ as a function of the top gate voltage $V_{tg}$ at various temperatures. For low values of the top-gate voltage $V_{tg}$, the conductance follows a thermally activated behavior and decreases with temperature. Above $V_{tg}$ ∼1-2 V, depending on the temperature, monolayer MoS$_2$ enters a metallic state, manifested by an increasing conductance as the temperature is decreased. **b,** Temperature dependence of the conductance for different values of charge density $n_{2D}$. **c,** Mobility dependence on temperature shows a mobility practically independent of temperature under 30K, indicating that the deposition of the top-gate dielectric has resulted in the screening of charged impurities. In this temperature range, the mobility of this monolayer MoS$_2$ is ∼160 cm$^2$/Vs. Above ∼100K the mobility decreases due to phonon scattering and follows a $T^\gamma$ dependence with $\gamma = 0.73$. The exponent $\gamma$ is strongly reduced with respect to its value for a single-gated device ($\gamma = 1.4$) and is indicative of phonon mode quenching due to the presence of the top-gate dielectric.

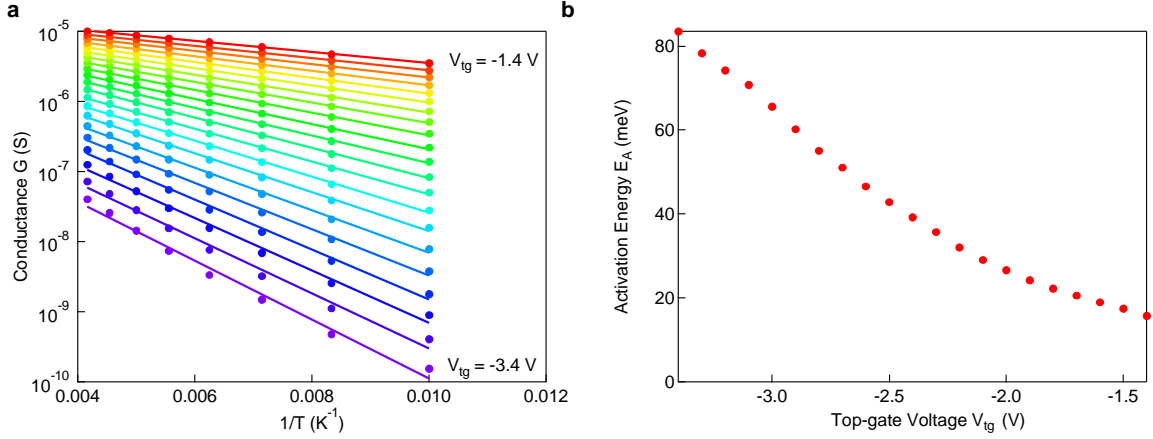

**Figure 4. Activation energies $E_a$ for monolayer MoS$_2$ in a top-gated configuration in the insulating regime. a,** Arrhenius plot of the conductance of monolayer MoS$_2$ covered with HfO$_2$, in the insulating regime. **b,** Dependence of activation energy $E_a$ on $V_{tg}$.

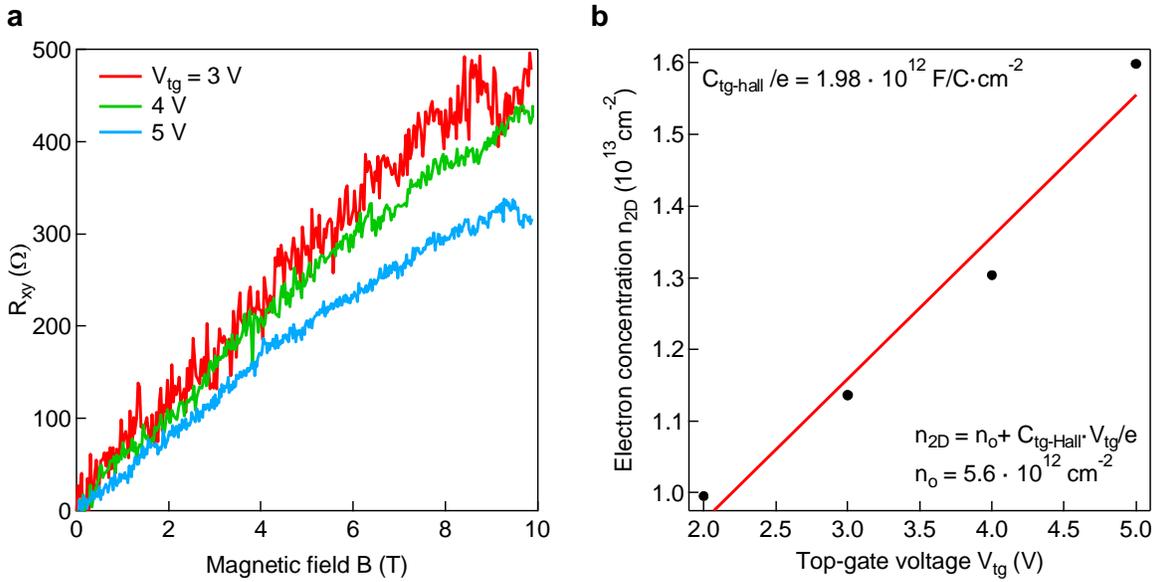

**Figure 5. Hall-effect measurements in dual-gated monolayer MoS$_2$ devices. a,** Hall resistance $R_{xy}$ versus magnetic field $B$ for different positive top-gate bias voltages $V_{tg}$. **b,** Electron concentration $n$ extracted from $R_{xy}$ for different values of the top-gate voltage $V_{tg}$. From the slope of the red solid line we calculate the capacitance per unit area $C_{tg,Hall}$ of the top-gate MoS$_2$ device. The residual doping of the MoS$_2$ channel is $n_o = 5.6 \cdot 10^{12}$ cm$^{-2}$. All measurements are performed at $T = 4$ K with a grounded back-gate electrode.

# Supporting information

# for

# Mobility engineering and metal-insulator transition in monolayer MoS$_2$

B. Radisavljevic and A. Kis[*]

*Electrical Engineering Institute, EcolePolytechniqueFederale de Lausanne (EPFL), CH-1015 Lausanne, Switzerland*

*Correspondence should be addressed to: Andras Kis, andras.kis@epfl.ch

**Device details**

We have performed measurements on two devices in single-gate configuration, two devices in single-gate configuration covered with a 30 nm thick $HfO_2$ layer and six devices in dual-gate configuration. Their characteristics are summarized in the following table:

**Table 1.** Device details

| Device | Configuration | W (μm) | $L_{12}$ (μm) | $k_F \cdot l_e$ | $n_{MIT}$ ($10^{13}$ cm$^{-2}$) | $\gamma$ | μ(cm$^2$/Vs) T=4K | μ(cm$^2$/Vs) T=260K |
|---|---|---|---|---|---|---|---|---|
| Monolayer1 | Single-gate | 3.9 | 0.7 | - | - | 1.4 | 6.4 | 17.2 |
| Monolayer2 | Dual-gate | 3.0 | 1.2 | 1.8 | 1 | 0.3 | 87.7 | 56.9 |
| Monolayer3 | Dual-gate | 3.7 | - | - | - | 1.29 | 50.1 | 16.5 |
| Monolayer4 | Dual-gate | 3.0 | 1.4 | 0.9 | - | 0.52 | 46.2 | 13.9 |
| Monolayer5 | Dual-gate | 1.9 | 1.6 | 2.5 | 1 | 0.73 | 160 | 63.7 |
| Monolayer6 | Dual-gate | 2.3 | 2.00 | 1.76 | - | 0.53 | 60 | 31.1 |
| Monolayer7 | Single-gate, with dielectric | 3.20 | 1.3 | - | - | - | 30.9 | 30.6 |
| Double-layer | Dual-gate | 1.6 | 1.6 | 2 | 1.3 | 1.47 | 117.6 | 26.4 |
| Three-layer | Single-gate, with dielectric | 4.9 | 1.8 | 2 | 1.1 | 0.75 | 84 | 24 |
| Four-layer | Single-gate | 4.6 | 1.3 | - | - | - | 1 | 30.1 |

W is the channel width and $L_{12}$ is the distance between voltage probes used in four-contact measurements. $k_F \cdot l_e$ is the Ioffe-Regel parameter related to the metal-insulator transition point and $n_{MIT}$ is the electron concentration at which the transition occurs, extracted from Hall-effect measurements.

**Capacitance determination**

Extract device capacitance from Hall effect measurements and the transverse Hall resistance $R_{xy}$ for all MoS$_2$ devices covered with a dielectric layer in order to accurately determine the mobility. The contact resistance for uncovered devices is too large to perform meaningful $R_{xy}$ measurements. From the inverse slope of $R_{xy}$ vs magnetic field (an example is shown on figure 5a in the main manuscript), we can directly determine the electron density $n_{2D}$ in the MoS$_2$ channel. The variation of the electron density extracted from $R_{xy}$ as a function of the control-gate voltage for two typical situations encountered in the literature is shown on Figure S1. On figure S1a, we show the dependence of the charge density on the back-gate voltage for a device in which the MoS$_2$ channel is covered with a 20nm thick HfO$_2$ layer. From the slope, we can extract the correct capacitance of the back-gate, $C_{bg\text{-}Hall}$, which in this case is 2.4 times higher than the capacitance calculated using the parallel-plate capacitance model $C_{geom} = \varepsilon_0 \varepsilon_r / d_{ox,bottom}$. The capacitive coupling between the MoS$_2$ channel and the back-gate is therefore increased due to the presence of the dielectric covering MoS$_2$ and any mobility estimate that would use the geometric capacitance instead of would yield a mobility value overestimated by a factor of 2.4. Similarly, in Figure S1B we present charge density measurements for a device in which the top gate has been disconnected. In this case we find that the capacitive coupling is increased by a factor of 53. This shows that using the parallel-plate capacitance model in place of an actual, measured capacitance in this type of situations can result in underestimating the strength of the capacitive coupling and field-induced charge density and lead to an overestimated mobility.

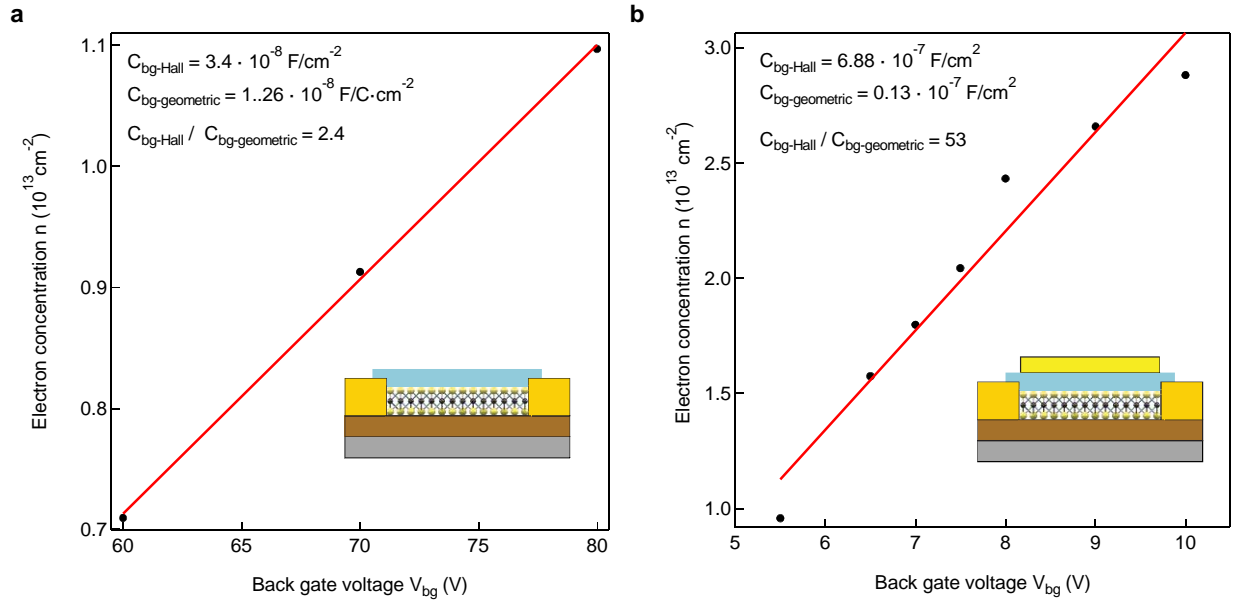

**Figure S1. Electron concentration *n* extracted from $R_{xy}$ for different values of the control gate voltage. a,** Charge density vs. bottom gate voltage for the three-layer device from Table 1. The conductivity is controlled using a bottom gate, while the channel is covered by a 30nm thick $HfO_2$ layer. The presence of the dielectric increases the back-gate capacitance by a factor of 2.4 with respect to the parallel-plate capacitance, commonly used for mobility estimates. **b,** Charge density vs. bottom gate voltage for the top-gated monolayer device (monolayer 4 in Table 1) measured as a function of the bottom gate while the top gate is disconnected. The capacitance is increased by a factor of 53 with respect to the parallel-plate capacitance where one plate is the back-gate and the other the $MoS_2$ channel.